
%
patchlevel 431.

\magnification = \magstep1
\overfullrule=0pt

\def\a{\alpha}
\def\Af {A^{(1)}}
\def\As{A^{(2)}}
\def\Bf {B^{(1)}}
\def\Bs{B^{(2)}}

\def\b{\beta}
\def\bull{\vrule height 1.5ex width 1.4ex depth -.1ex}

\def\ca{{\cal A}}
\def\cc{{\cal C} }
\def\cf{{\cal F}}
\def\cg{{\cal G}}
\def\ct{{\cal T}}

\def\d{\delta}
\def\D{\Delta}

\def\e{\epsilon}

\def\la{\langle}

\def\lp{(L \times L)'\times \ovg/L}
\def\lpw{\bigl(\lp\bigr)/W}

\def\o{\omega}
 \def\O{\Omega}
\def\ova{{\ov a}}
\def\ovb{{\ov b}}
\def\ovy{{\ov y}}
\def\ov{\overline}
\def\ovg{{\ov G}}
\def\ovx{{\overline x}}
\def\ovy{{\ov y}}

\def\ra{\rangle}

\def\S{\Sigma}

\def\pr{\partial}

 \def\to{\rightarrow}
\def\th{\theta}
\def\Th{\Theta}
\def\tl{\tilde}

\def\ug{{\underline g} }

\def\RefHead{\vglue 24pt \hfil {\bf REFERENCES} \hfil
\vglue 12pt}

\def\References#1#2{
{ \advance\leftskip by 1in
\noindent
\llap{\hbox to .5in{[#1] \hfil}}#2.

}}

\def\Title#1#2{\vglue .5in
\centerline{\bf   #1} \vskip .5pc
\centerline{\bf  #2} \vskip 1.25pc}

\def\Author#1{\vskip .5pc
\centerline{#1}}
\def\Address#1{
\centerline{\it #1}  }

\def\Abstract#1{\vglue .3in
{\parindent=.7in
\narrower
{\noindent{\bf Abstract.} \enskip #1}

}
\vglue .5in}

\Title{The Semiclassical Limit }
{for $SU(2)$ and $SO(3)$ Gauge Theory on the Torus}
\Author{Ambar Sengupta\footnote{}{
Research supported in part by  LEQSF Grant RD-A-08.}}
\Address{Department of Mathematics}
\Address{Louisiana State University}
\Address{Baton Rouge, Louisiana 70803-4918}
\Address{e-mail : sengupta@marais.math.lsu.edu}
\vglue 12pt

\Abstract{ We prove that  for $SU(2)$ and $SO(3)$  quantum gauge
theory on a
torus, holonomy expectation values with respect to the Yang-Mills
measure
$d\mu_T(\o) =N_T^{-1}e^{-S_{YM}(\o)/T}[{\cal D}\o]$ converge, as
$T\downarrow
0$, to integrals with respect to a   symplectic volume measure
$\mu_0$ on the
moduli space of flat   connections on the bundle. These moduli spaces
and the
symplectic structures  are described explicitly.}

\vskip 0.5in

\noindent{\bf 1. Introduction}
\vskip 0.10in

In this paper we  prove  that for $SU(2)$ and $SO(3)$  quantum gauge
theory on
a torus,  the holonomy (`Wilson loop')  expectation values for the
Yang-Mills
measure $d\mu_T(\o) =N_T^{-1}e^{-S_{YM}(\o)/T}[{\cal D}\o]$ (the
notation is
explained in  subsection 2.5 below) converge, as $T\downarrow 0$, to
integrals
with respect to a   symplectic volume measure $\mu_0$ on the moduli
space of
flat   connections on the bundle.   We  also show  that   for the
non-trivial
$SO(3)-$bundle over the torus, the moduli space of flat connections
consists of
just one point and the limiting measure exists  and is thus, of
course, just
the unit mass on this point (a similar  situation exists in genus
$0$, which is
treated from a slightly different point of view in   [Se 1]). The
proofs are by
direct computation using the expectation value formulas derived from
a
continuum quantum gauge theory in [Se 2,3] and by lattice theory in a
number of
works including [Wi 1] (the work [Wi 2] also contains results of
related
interest), and the description of the symplectic form obtained in [KS
1].

The most significant  result related to the present work is the
corresponding
result by Forman [Fo] for gauge theory on compact orientable surfaces
of genus
$>1$. Forman's proof relies on results of Witten [Wi 1]; a  more
direct proof
of  part of  Forman's  result  has  been obtained  by C. King and the
author in
[KS 2]. The main case we work with in this paper, genus $1$ and gauge
group
$SU(2)$,  is singular in two ways (thereby making the method used in
[Fo],
[Wi], [KS 2] inapplicable) : (1) the `partition function' goes to
$\infty$, as
$T\downarrow 0$, and (2) no flat connection is  irreducible.  The
situation
over the torus is singular   for other gauge groups  as well, but
the case
$SU(2)$ (or $SO(3)$) may   deserve this separate study because  the
symplectic
structure on the space of flat connections is, in this case,  very
simple and
thus a  completely `hands-on'  study is possible.

\vskip .25in
\noindent{\bf  2. The Limiting Expectation Values}.
\vskip 0.10in

{\it  2.1. The torus as a quotient of the disk} : Let $\S$ be a torus
equipped
with a Riemannian metric, scaled so that the total area of  $\S$ is
$1$.  The
area of $A\subset \S$ will be denoted $|A|$. As a manifold, $\S$ may
be
obtained from the closed planar disk $D=\{(x,y)\in R^2 :x^2+y^2\leq
1\}$,
centered at $O=(0,0)$, as follows. Let $x_t=\bigl(  \cos(2\pi t),
\sin(2\pi
t)\bigr)$, for $t\in R$, and let $K_i$ be the arc $[{{i-1}\over
4},{i\over
4}]\to\pr D: t\mapsto x_t$. Thus  the $K_i$ ($i=1,2,3,4$) split $\pr
D$  into
four congruent arcs. Identify  $K_1$ with $K_3$ reversed, and $K_2$
with $K_4$
reversed, linearly; i.e. $x_t$ is identified with $x_{t'}$ whenever
$t-{{i-1}\over 4}={{i+2}\over 4}-t'$ with $i\in\{1,2\}$, $t\in
[{{i-1}\over
4},{i\over 4}]$ and $t'\in [{{i+1}\over 4},{{i+2}\over 4}]$. The
quotient space
is a torus which we take to be $\S$, and we denote the quotient map
by
$q:D\to\S$.  The two  loops $S_1=q(x_0O).q(K_1).q(Ox_0)$ and
$S_2=q(x_0O).q(K_2).q(Ox_0)$,  where $Ox_0$ is radial, generate the
fundamental
group $\pi_1\bigl(\S, q(O)\bigr)$, with ${\ov S}_2{\ov S}_1S_2S_1$
being
homotopic to the constant curve at $q(O)$. For technical convenience,
we equip
$\S$ with the orientation induced by $q$ from the standard
orientation of
$D\subset R^2$.

{\it 2.2. Triangulation of $\S$,  and Lassos} : We will work with a
fixed
triangulation  $\cal T$ of $\S$.  In order that we can apply some of
the
results of [Se 2,3] it is necessary to put some restrictions, which
we describe
below,  on $\cal T$.  We will not make  any significant  overt   use
of these
restrictions, and the reader may choose to proceed by viewing Theorem
2.10 as a
`definition'. In any case, every triangulation of $\S$ has a
subdivision which
is isomorphic to a subdivision of  a triangulation which satisfies
the
restrictions.   The requirement on the  triangulation of $\S$ is that
it be the
projection by $q$ of a triangulation of $D$ which satisfies  : (i)
each
oriented $1-$simplex  is either radial or `cross-radial' (i.e.
intersects each
radius at most once),  (ii)  each $K_i$   is the composite  of
$1-$simplices,
and (iii) each   $0-$simplex  is an endpoint of a $1-$simplex which
is part  of
a sequence of radial $1-$simplices going from $O$ to a point on $\pr
D$ (in
particular, $q(O)$ is a $0-$simplex of the triangulation).   If $\D$
is an
oriented $2-$simplex of $\ct$ then there is a radial path $Ox$ in $D$
whose
projection by $q$ is a path $B$, consisting of oriented
$1-$simplices, from
$q(O)$ to a $0-$simplex on $\pr\D$.  The corresponding loop ${\ov
B}.\pr \D.B$
will be called a  {\it  lasso}. We will let $\{\D_1,...,\D_n\}$ be
the
positively oriented two-simplices of the triangulation $\cal T$  of
$\S$, and
${\cal E}=\{e_1,{\ov e}_1,...,e_M, {\ov e}_M\}$   the set  of
oriented
one-simplices of the triangulation (wherein ${\ov e}_i$ is $e_i$ with
the
opposite orientation).   We will often work with loops $C_1,...,C_k$
which are
composites of $1-$simplices.

{\it 2.3. The groups $G$ and $\ovg$, the Lie algebra $\ug$, and the
metric
$\la\cdot,\cdot\ra_\ug$} : All through this paper  $G$ will denote
the group
$SU(2)$, while $\ovg$ will denote either $SU(2)$ or $SO(3)$.  Thus
there is a
(universal) covering map   $G\to \ovg$ (which is the identity map if
$\ovg=G$).
 The Lie  algebra $\ug$ of $G$ will be equipped with a fixed
$Ad-$invariant
metric $\la\cdot,\cdot\ra_\ug$. We will write $I$ to denote both the
identity
element in $G$ and the identity element in $\ovg$.

{\it  2.4. Principal $\ovg-$bundle} : Let $\pi:P\to\S$  be a
principal
$\ovg-$bundle over $\S$.
This is classified, relative to the fixed orientation on $\S$, up to
bundle
equivalence  by an element $h\in ker (G\to \ovg)$ (one may take
equation (3.1)
of Section 3 as defining $h$). Here this kernel  is either $\{I\}$ or
$\{\pm
I\}$ ($I$ being the $2\times 2$ identity matrix); if  $h=I$ then the
bundle is
trivial, while if $h=-I$ (which is possible only when $\ovg =SO(3)$)
the
bundle is non-trivial.

{\it  2.5. The Yang-Mills measure on the space $\cc$} :  Let $\ca$ be
the set
of all connections on $P$ and let $\cg_m$ be the group of all bundle
automorphisms of $P$, covering the identity map $\S\to\S$, and fixing
the fiber
$\pi^{-1}\bigl(m\bigr)$ pointwise, where $m=q(O)$.  The group $\cg_m$
acts on
$\ca$ by pullbacks of connections : $(\phi,\o)\mapsto \phi^*\o$. The
Yang-Mills
measure $\mu_T$ is a probability measure on  a certain `completion'
$\cal C$
of the quotient space $\ca/\cg_m$.  The construction of $\mu_T$ is
carried out
in [Se 2,3].  Formally, $d\mu_T(\o) =N_T^{-1} e^{-S_{YM}(\o)/T}
[{\cal D}\o]$,
where $S_{YM}$ is the Yang-Mills action functional, $[{\cal D}\o]$ is
the
pushforward on $\cal C$ of   ``Lebesgue measure'' on $\ca$, and $N_T$
is a
``normalization constant''.

{\it  2.6. Stochastic holonomy} :  Fix once and for all, a point $u$
on the
fiber $\pi^{-1} (m) $, where $m=q(O)$.  If $C$ is a piecewise smooth
closed
curve on $\S$, based at $q(O)$, then corresponding to a connection
$\o$ on $P$
there is associated the {\it holonomy} $g_u(C;\o)\in \ovg$ of $\o$
around $C$
with initial point $u$.  The holonomy $g_u(C;\o)$ remains invariant
if $\o$ is
replaced by $\phi^*\o$, for any $\phi\in\cg_m$.  If $C$ is a closed
curve on
$\S$, based at $q(O)$,
which is the composite of  oriented $1-$simplices of $\ct$, then (as
shown in
[Se 2,3])  there is a $\mu_T-$almost-everywhere defined measurable
function
$\o\mapsto g(C;\o)\in\ovg$ on $\cc$.
 This function may be called the `stochastic holonomy around $C$'
and is
defined by reinterpreting the classical equation of parallel
transport as a
stochastic differential equation.  If $C_1,...,C_k$ are $k$ such
closed curves,
and $f$ is a bounded measurable function  on $G^k$, then the
expectation value
$$\int_\cc f\bigl(g_u(C_1;\o),...,g_u(C_k;\o)\bigr)\,d\mu_T(\o)$$
is  of interest.  We state in Theorem 2.10 below an explicit formula
for   the
expectation value.
Part of the goal of this paper is the determination of  the limit of
this
expectation value, for continuous $f$, as $T\downarrow 0$.

{\it  2.7. Notation} ($b\mapsto x_b$)  : As before, let
$\{\D_1,...,\D_n\}$ be
the positively oriented two-simplices of the triangulation $\cal T$
of $\S$,
and  ${\cal E}=\{e_1,{\ov e}_1,...,e_M, {\ov e}_M\}$   the set  of
oriented
one-simplices of the triangulation.  We shall use  maps ${\cal E}\to
{  G}
:b\mapsto {  x}_b$ for which ${ x}_{\ov b} = { x}_b^{-1}$, where
${\ov b}$
denotes the orientation-reverse of the oriented $1-$simplex  $b$. If
$C$ is a
closed curve which is the composite  $b_{r }...b_{ 1}$ of oriented
$1-$simplices  then we write ${  x}(C)$ to mean ${  x}_{b_{r }}...{
x}_{b_{
1}}$. Thus ${  x}(\pr \D_i)\in  G$, where $\pr \D_i$ is the boundary
of $\D_i$
with some choice of initial point. We write $\ovx_b$ for the element
of $\ovg$
covered by ${  x}_b\in { G}$; so $\ovx (C)\in G$ is covered by ${
x}(C)\in  G$.

{\it  2.8. The Brownian Density (or `Heat Kernel')  $Q_t(x)$ } :  Let
$Q_t(x)$
be the density, with respect to the Haar measure $dx$ on $G$  (of
total mass
$1$), at time $t (>0)$ of standard Brownian motion on $G$ (driven by
the
$Ad-$invariant metric $\la\cdot,\cdot\ra_\ug$ on $\ug$).

{\it 2.9. Facts about $Q_t(x)$} :  $Q_t(x)$ is a multiple of  the
fundamental
solution  of the heat equation  on $G$. It can be expressed as
$Q_t(x)
=\sum_{n=1}^\infty e^{-C_nt/2}n\chi_n(x)$, where $\chi_n$ is the
character of
the $n-$dimensional irreducible representation of $G$, and
$C_n=(n^2-1)/\kappa^2$ (wherein $\kappa$, specifying the metric
$\la\cdot,\cdot\ra_\ug$, is the   length of the vector  $
\left(\matrix{  i &
0   \cr
 0 &  -i\cr}\right) \in\ug$),
the series being absolutely convergent.  For any $\d>0$,
$\lim_{t\downarrow
0}\sup_{|x-I|>\d}Q_t(x)=0$, wherein $|x-I|$ denotes the distance
between $x$
and $I$. If $f$ is a bounded measurable function on $G$ and is
continuous near
$I$,  then $\lim_{t\downarrow 0}\int_G f(x)Q_t(x)\,dx=f(I)$.

{\it  2.10. Theorem } ([Se 2,3]) :  Let $\{\D_1,...,\D_n\}$ be the
positively
oriented two-simplices of the triangulation $\cal T$  of $\S$, and
${\cal
E}=\{e_1,{\ov e}_1,...,e_M, {\ov e}_M\}$   the set  of oriented
one-simplices
of the triangulation. Suppose   that $ C_1,...,C_k $ are loops on
$\S$, all
based at $m$ (which, recall,  is assumed to be a $0-$simplex of the
triangulation), which are   composites of oriented $1-$simplices.
Then :
$$\displaylines{ \int_{{ \cc}}
f\bigl(g(C_1;\o),...,g(C_k;\o)\bigr)\,d\mu_T(\o)
=\cr
\hfill  {1\over {Z^h_T}}\int_{{ G}^M}f\bigl({\ov x}(C_1),...,{\ov
x}(C_k)\bigr)
{  Q}_{T|\D_1|}\bigl(h{  x}(\pr \D_1)\bigr)\prod_{i=2}^n
{Q}_{T|\D_i|}\bigl({
x}(\pr \D_i)\bigr) \,d{  x}_{e_1}...d{  x}_{e_M}}$$

  where
$$Z^h_T = \int_{{  G}^{2 }} {  Q}_{T}(hb^{-1}a^{-1}ba)\,da.db.\quad
\bull$$

{\it 2.11.  The variables $y_{\D_i}$ and the function $F$} : Given
$b\mapsto
x_b$, as above, it is possible, with an appropriate indexing of the
$2-$simplices as $\D_1,...,\D_n$,  to introduce   $y_{\D_i}\in G$,
one for each
oriented $2-$simplex $\D_i$, and $a,b\in G$, such that  : (i)
$y_{\D_i}
=x(B)^{-1} x(\pr \D_i) x(B)$ for  the lassos ${\ov B}.\pr\D_i.B$
(for
appropriate choices of $B$), $i\in\{2,...,n\}$,    while  for  the
remaining
simplex $\D_1$ :   $y_{\D_1} =
hx(B)^{-1}x(\D_1)x(B)=hb^{-1}a^{-1}ba(y_{\D_2}...y_{\D_n})^{-1}$,
(ii)
$x(S_1)=a$, and $x(S_2)=b$, and (iii)  each $x(C_i)$ is a product of
suitable
powers of the $y_{\D_i}$ and $a,b$. The latter actually follows from
(i) and
(ii)  using the fact that every loop of the type $C_i$ can be
obtained from the
composite of  lassos and reversed lassos and the basic loops $S_1$
and $S_2$
(and their reverses) by dropping certain $1-$simplices which are
traversed in
opposite directions consecutively.   These facts are proven in  [Se
2] (Lemmas
A2 and A3).  Thus there is a function $F$ such that

$$f\bigl({\ov x}(C_1),...,{\ov x}(C_k)\bigr) = F(\{{\ov y}_\D\},{\ov
a},{\ov
b})  \eqno(2.1)$$

wherein we have written  $ F(\{{\ov y}_\D\},{\ov a},{\ov b}) $ to
mean
$F({\ovy}_{\D_2},...,{\ovy}_{\D_n},\ova,\ovb)$.

Thus $ F(\{{\ov y}_\D\},{\ov a},{\ov b})$ is obtained from
$f\bigl(\ovx(C_1),...,\ovx(C_k)\bigr) $ by writing each $\ovx(C_i)$
as a
product of powers of  the $\ovy_{\D_i}$ and $\ova,\ovb$. In
particular,   if
$f$  is continuous then so is $F$.   Moreover,

$$\displaylines{\int_{{ G}^M}f\bigl({\ov x}(C_1),...,{\ov
x}(C_k)\bigr)
Q_{T|\D_1|}\bigl(h  x(\pr \D_1)\bigr)\prod_{i=2}^n {
Q}_{T|\D_i|}\bigl({
x}(\pr \D_i)\bigr) \,d{  x}_{e_1}...d{  x}_{e_M}\cr
\hfill = \int  F(\{{\ov y}_\D\},{\ov a},{\ov b})Q_{T|\D_1|}\bigl(h
b^{-1}a^{-1}ba(y_{\D_2}...y_{\D_n})^{-1}\bigr)\prod_{i=2}^nQ_{T|\D_i|}
(y_{\D_i})\,dadb.dy_{\D_2}...dy_{\D_n} \quad\quad(2.2)}$$

This is  proven  in (Lemma 8.5 of) [Se 3] (and in [Se 2] in the case
$h=I$).

{\it 2.12. Remark.}  If $\o$ is a connection on $P$, then by setting
${\ov
y}_{\Delta_i}=g_u({\ov B}.\pr \D_i . B;\o)$ (notation as in
subsection  $2.11$
above), and  $a=g_u(S_1;\o)$, $b=g_u(S_2;\o)$, we can take $x_b  $ to
be the
element of $G$ describing  $\o-$parallel transport along  $b$ (as
measured by a
suitable section of $P$ over the bonds of the triangulation). Then
$x(C_i)=g_u(C_i;\o)$. In particular, if $\o$ is flat, then equation
$(2.1)$
implies that :
$$f\bigl(g_u(C_1;\o),...,g_u(C_k;\o)\bigr) = F(\{e\},
g_u(S_1;\o),g_u(S_2;\o)\bigr) \eqno(2.1')$$

{\it 2.13. Lemma}.  For $T>0$, we have :
$$Z^h_T=\sum_{n=1}^\infty e^{-C_nT/2}   {{\chi_n(h)}\over {n}},$$
where $C_n= (n^2 -1)/\kappa^2$ (wherein $\kappa>0$ specifies the
metric on
$\ug$, as in Facts 2.9), and ${{\chi_n(h)}\over {n}}$ is $1$ if $h=I$
(i.e.
when $P$ is trivial) and is $(-1)^{n+1}$ when $h=-I$ (i.e. when
$\ovg=SO(3)$
and the bundle is non-trivial). In particular, the series above is
convergent.
Moreover,

$$\lim_{T\downarrow 0}Z_T^h =\cases{\infty & if $h=I$;\cr
{1\over 2} &if  $ h=-I$.\cr}$$

{\it Proof} : With $\chi_n$ as in Facts 2.9,
$\int_G\chi_n(hy^{-1}x^{-1}yx)\,dxdy={1\over {n^2}}\chi_n(h)$ (see
Ex. 2.4.17.3
and Proposition 2.4.16 (iii)  in  [BrtD]), where the integration is
with
respect to Haar measure of total mass $1$.   The expression for
$Z^h_T$
follows upon integration term-by-term of the series   for $Q_T(h
y^{-1}x^{-1}yx)$  given in Facts 2.9. Term-by-term integration is
valid by
dominated convergence, since  (see Facts 2.9),
 $$\sum_{n=1}^\infty e^{-C_nT/2}n\int_G|\chi_n(h
y^{-1}x^{-1}yx)|\,dxdy\leq
\sum_{n=1}^\infty e^{-C_nT/2}n.n=Q_T(I)<\infty.$$
  Next, if $h=I$, we see by monotone convergence  that
$Z^I_T=\sum_{n=1}^\infty e^{-C_nT/2}\to\infty$, as $T\downarrow 0$.

To evaluate      $\lim_{T\downarrow 0}Z^{-I}_T$ we use the theta
function
identity :
$$1+2\sum_{n=1}^\infty \cos(2\pi nx) e^{- \pi n^2 t/2} =
t^{-1/2}\sum_{m\in
{\bf Z}} e^{-2\pi (x-m)^2/t},$$
wherein ${\bf Z}$ is the set of all integers.
Setting $x=1/2$ we obtain :
$$\sum_{n=1}^\infty (-1)^{n+1}e^{-\pi n^2 t/2} ={1\over 2} -{1\over
2}t^{-1/2}
\sum_{m\in {\bf Z}}e^{- 2 \pi (m-{1\over 2})^2/t}.$$
The left side of this equation is $e^{-\pi t/2}Z^{-I}_{\pi
t\kappa^2}$.
Letting $t\downarrow 0$, we obtain $\lim_{T\downarrow 0}Z^{-I}_T
={1\over 2}.$

In  Section 3 (following Theorem  3.15) we will calculate
$\lim_{T\downarrow
0}Z^{-I}_T ={1\over 2}$ by a different method. $\bull$

{\it  2.14. Lemma}. Let  $C_1,...,C_k$ be a collection of loops  as
above,
$f$ a continuous function on $G^k$, and let $F$ be the function
described in
equation (2.1). Then :

$$\lim_{T\downarrow 0} \int_{{ \cc}}
f\bigl(g(C_1;\o),...,g(C_k;\o)\bigr)\,d\mu_T(\o)=\lim_{T\downarrow
0}{1\over
{Z^h_T}}\int F(\{{\ov
y}_\D=I\},\ova,\ovb)Q_T(hb^{-1}a^{-1}ba)\,dadb$$

provided the limit on the right hand side exists.

{\it Proof} :   By continuity,  $ F(\{{\ov y}_\D\},{\ov a},{\ov b})$
is
uniformly close to $F(\{I\},{\ov a},{\ov b})$ when the $y_{\D_i}$ are
close to
$I$. Let $U_\d =\{y_{\D_i}:|y_{\D_i}-I|>\d,\,i=2,...,n\}$, for any
$\d>0$,
wherein $|y_{\D_i}-I|$ denotes the distance between $y_{\D_i}$ and
$I$.  In
view of the expectation value fromula given in Theorem 2.10 and
equation
(2.2),  it will suffice to show that, for any $\d>0$,
$${1\over
{Z^h_T}}\int_{U_\d}Q_{T|\D_1|}\bigl(hb^{-1}a^{-1}ba(y_{\D_2}...y_{\D_n
})^{-1}\bigr)\prod_{i=2}^nQ_{T|\D_i|}(y_{\D_i})\,dadb.dy_{\D_2}...dy_{
\D_n} $$
 converges to $0$, as $T\downarrow 0$.

The integral appearing above is dominated by
$$ \prod_{i=2}^n  \sup_{|y_i-I|>\d}Q_{T|\D_i|}(y_i),$$

and by Facts 2.9, this goes to $0$, as $T\downarrow 0$. The desired
result now
follows by taking into account Lemma 2.13. $\bull$

{\it  2.15. Remark}.    In view of Theorem 2.10, equations (2.1) and
(2.2),
and Remark 2.12, we see that  Lemma 2.14 says essentially that
$\lim_{T\downarrow 0}\mu_T$, if it exists in a  suitable weak sense,
{\it lives
on the space of   flat connections} on $P$.

{\it 2.16. Remark}.   Lemma 2.14 also shows that the determination of
the
``weak limit'' of  $\mu_T$, as $T\downarrow 0$,  is essentially
reduced to  the
determination of the limit
$$\lim_{T\downarrow 0} {1\over {Z_T^h}}\int_{G^2}  Fs({\ov a}, {\ov
b})
Q_T(haba^{-1}b^{-1})\,da.db,$$

for continuous  functions $F$ on ${\ov G}^2$.

{\it 2.17. Lemma}. If $F$ is continuous on $G^2$ then
$$\lim_{T\downarrow 0}{1\over {Z^I_T}}\int_{G^2} F(a,b)Q_T(
b^{-1}a^{-1}ba)\,da
db =\lim_{n\to\infty} n.\int_{G^2}F(a,b)\chi_n(
b^{-1}a^{-1}ba)\,da.db,$$
provided the limit on the  right hand side exists (as a finite
number).

{\it Proof} :  This is an application  of   the L'Hopital  rule.
Using Facts
2.9, we have  as in the proof of Lemma 2.13 :

$$\int_{G^2} F(a,b)Q_T( b^{-1}a^{-1}ba)\,da db =\sum_{n=1}^\infty
e^{-TC_n/2}\a_n,$$
where
$$\a_n =n.\int_{G^2} F(a,b)\chi_n(b^{-1}a^{-1}ba)\,da.db.$$
Taking $F$ to be $1$, we have, as in  Lemma 2.13,
$$Z^I_T =\sum_{n=1}^\infty e^{-C_nT/2}.$$

Assume that $\a_n\to L$, a finite number, as $n\to\infty$.  Let
$\e>0$, and
choose $N$ so that $|\a_n-L|<\e$ when $n\geq N$. Then :
$$\eqalign{{1\over {Z^I_T}}\int_{G^2} F(a,b)Q_T( b^{-1}a^{-1}ba) \,da
db
&={{\sum_{n<N}e^{- C_nT/2}\a_n}\over {Z^I_T}} + {{\sum_{n\geq N}
e^{-C_nT/2}\a_n}\over {\sum_{n\geq N}e^{-C_nT/2}}}\Bigl\{1-{
{\sum_{n<N}e^{-
C_nT/2} }\over {Z^I_T} }\Bigr\}\cr
&=O(T) + (L+\e')\bigl(1-O(T)\bigr),\quad\hbox{where}\,\,
0<\e'<\e,\cr}$$

where we have used Lemma 2.13 in the second equality.

So
$$|{1\over {Z^I_T}}\int_{G^2} F(a,b)Q_T( b^{-1}a^{-1}ba)\,da.db -L|
\leq  O(T)
+L.O(T) +\e. \bigl(1-O(T)\bigr).$$

The  required  result follows  upon taking $\limsup$ as  $T\downarrow
0$,
since  $\e>0$ is arbitrary. $\bull$

{\it 2.18. Notation $(k_t,a_\th,K)$/Some facts about $SU(2)$. }
Every element
$g\in SU(2)$ can be written in the form :
$$g = k_\phi a_\th k_\psi\eqno(2.3)$$

with $0\leq \phi < 2\pi$, $0< \th < \pi$, $0 \leq \psi < 2\pi$ (these
correspond to  the `Euler angles' for $SO(3)$),   where

$$k_t = k(e^{it}) =  \left(\matrix{e^{i{t}}&0\cr
0& e^{-i{t}}\cr}\right)$$

and

$$a_\th = a(\th) =\left(\matrix{  \cos{\th\over 2}& i\,
\sin{\th\over 2}\cr
i\,   \sin{\th\over 2}&   \cos{\th\over 2}\cr}\right).$$

We denote by  $K$ the maximal torus $\{k_t :t\in {\bf  R}\}$  in $G$.

Moreover,  the  Haar measure $dg$ on $G$ of unit total  mass can be
expressed
as :

$$dg = {1\over { 8\pi^2}}   \sin\,\th\, d\phi   d\th d\psi .$$

To be precise, the map $(0,2\pi)\times(0,\pi)\times (0,2\pi)\to G :
(\phi,\th,\psi)\mapsto  k_\phi a_\th k_\psi$ is a two-to-one smooth
local
diffeomorphism onto  a dense open subset of $SU(2)$, with Jacobian  $
{1\over {
4\pi^2}}   \sin\,\th$ (the expression for $dg$ has a further factor
of ${1\over
2}$ because of the two-to-one nature of   $ (\phi,\th,\psi)\mapsto
k_\phi
a_\th k_\psi$).

If $H$ is any bounded measurable function on $SU(2)$ which is central
(in the
sense that $H(gxg^{-1}) = H(x)$ for every $x,g\in SU(2)$) then :

$$\int_{SU(2)} H(g)\,dg =  {2\over { \pi}}\int_0^{  \pi} dt\,
\sin^2 t.    H(
k_t ). $$
The reader may consult [BrtD] (1.5.20.6 and 2.5.2)  or [Waw] for
these  facts.
 $\bull$

{\it 2.19.  Proposition. } If $F$ is any smooth function on
$SU(2)\times
SU(2)$, then :

$$\lim_{n\to\infty} n.\int _{SU(2)\times SU(2)} F(a,b)\chi_n(
b^{-1}a^{-1}ba)\,da.db =\int_{K\times K} {\ov F}(k,k')\,dk.dk'
\eqno(2.5)$$
where $dk$ and $dk'$ are the Haar measure of unit total mass on  $K$,
and ${\ov
F}(a,b)$ is  the `average'  $\int_{SU(2)} F(gag^{-1},gbg^{-1})\,dg$.

{\it Proof} : Since the integral on the left in equation $(2.5)$ is
unaltered
if $F$ is replaced by ${\ov F}$, it will   suffice to assume   that
$F$, in
addition to being  smooth, is  a central function, i.e. that
$F(gag^{-1},gbg^{-1})=F(a,b)$ for every $g\in SU(2)$.  Denote by
$I_n$ the
integral on the left of  equation $(2.5)$.  Since  $b\mapsto
\int_{SU(2)}
F(a,b)\chi_n(b^{-1}a^{-1}ba)\,da$ is central,   we have  by Facts
2.18 :

$$nI_n ={{2n}\over \pi}\int_0^\pi dt\,
\sin^2t.\int_{[0,2\pi]\times[0,\pi]\times [0,2\pi]}{{d\phi.
\sin(\th)d\th.d\psi}\over {8\pi^2}}\, F(k_\phi a_\th
k_\psi,k_t)\chi_n( k_t
a_\th k_{-t} a_{-\th}) . $$

Now the character $\chi_n$ is given by :
$$\chi_n(g) = {{  \sin \bigl[n.   \cos^{-1}\{{1\over
2}Tr(g)\}\bigr]}\over {
\sin \bigl[   \cos^{-1}\{{1\over 2}Tr(g)\}\bigr]}}\quad\hbox{for
every}\,\,
g\in SU(2)\setminus\{\pm I\}$$

where $Tr$ denotes trace of a matrix, and  the same  range of $
\cos^{-1}$ is
used in numerator and denominator.  Calculation shows that :
$$ A(\th,t)\buildrel\rm def\over ={1\over 2}Tr(k_t a_\th k_{-t}
a_{-\th}) = 1 -
2 \bigl(\sin^2{\th\over 2}\bigr).   \sin^2t.$$

{}From  these formulas  we have the following useful relationship
(which is
verified by working out the derivative in the right side) :

$$n.  \sin^2(t).\sin \th .  \chi_n (k_t a_\th k_{-t} a_{-\th}) = -
{{\pr}\over
{\pr \th }}   \cos \Bigl[ n.   \cos^{-1}\bigl\{A(\th,t)\}\Bigr].$$

In view of  this,  integration by parts yields :

$$nI_n
 = B.T. + {2\over \pi}\int_0^\pi dt\,
\int_{[0,2\pi]\times[0,\pi]\times
[0,2\pi]} {{d\phi.   d\th.d\psi}\over {8\pi^2}}\, {{\pr F(k_\phi
a_\th k_\psi,
k_t)}\over {\pr\th}}   \cos \Bigl[ n.   \cos^{-1}A(\th,t)\Bigr]
\quad (2.6)$$

For the boundary term  $B.T.$ we have  :
$$\eqalign{B.T.  &=  -{2\over\pi}\int_0^\pi
dt\int_{[0,2\pi]^2}{{d\phi.d\psi}\over {8\pi^2}}\Bigl\{F(k_\phi a_\pi
k_\psi,
k_t)   \cos(2nt) - F(k_{\phi +\psi},k_t)\bigr\}\cr
&\rightarrow {2\over\pi}\int_0^\pi
dt\int_{[0,2\pi]^2}{{d\phi.d\psi}\over
{8\pi^2}}F(k_{\phi+\psi},k_t),\quad\hbox{as}\,\,n\to\infty,\,\hbox{(Ri
emann-Lebesgue lemma)}\cr
&=\int_{K}dk\int_0^\pi {{dt}\over {\pi}} F(k,k_t)\cr
&=\int_{K\times K} F(k,k')\,dk.dk'\cr}$$
where in the last step we used the conjugation-invariance property of
$F$ and,
for instance, $a_\pi k_ta_\pi^{-1}=k_{-t}$.

The rest of the argument is to show that the second term on the right
in
equation $(2.6)$ goes to zero as $n\to\infty$. To this end it will
suffice to
show that the integral
$$\int_0^\pi dt.\int_0^{2\pi}d\th. H'(t,\th).   \cos\bigl[n.
\cos^{-1}
A(\th,t)\bigr]$$
goes to $0$ as $n\to\infty$, for every continuous function $H'$ on
$[0,\pi]\times[0,2\pi]$.  The second factor in the integrand above is
invariant
under $t\mapsto\pi -t$ and $\th\mapsto 2\pi -\th$; thus  we can
reduce the
integral to the form :

$$J_n\buildrel \rm def\over = \int_{0}^{\pi/2}dt	\,\int_0^\pi
d\th \,\,
H(t,\th).   \cos\bigl[ n.  \cos^{-1}A(t,\th)\bigr]$$
where $H$ is continuous on $[0,\pi/2]\times[0,\pi]$. Thus, we wish to
show that
$J_n\to 0$, as $n\to\infty$. It will be convenient and sufficient to
assume
that $H$ vanishes near the boundary of $[0,\pi/2]\times[0,\pi]$.

For $t\in [0,{\pi\over 2}]$ and $\th\in [0,{\pi}]$, introduce $\xi\in
[0,{\th\over 2}]\subset [0,{\pi\over 2}]$ by :

$$  \sin\xi  =   \sin t .    \sin (\th/2)$$

Then $(t,\th)\mapsto (\xi,\th)$ is a diffeomorphism of  the open set
$(0,\pi/2)\times (0,\pi)$ onto  the open set  $\{(\xi,\th):
0<\xi<{\th\over
2}<{\pi\over 2}\}$, and the inverse map is of the form
$(\xi,\th)\mapsto\bigl(t(\xi,\th),\th\bigr)$.

We have

$$  \cos\xi \,d\xi \wedge d\th =   d(\sin\xi)\wedge d\th =\cos t
\sin
(\th/2)\, dt \wedge d\th = \sqrt{   \sin^2 (\th/2) -   \sin^2 \xi}
\,\,dt
\wedge d\th$$
where the positive square-root is taken because $  \cos t\geq 0$ and
$   \sin
{\th\over 2} \geq 0$  since $t,{\th\over 2}\in [0,\pi/2]$.  Thus

$$J_n =   \int_{0}^{\pi/2}d\xi  \Bigl\{\int_{2\xi}^\pi d\th\, {\tl
H}(\xi,\th)\Bigr\}.\cos (2n\xi)$$
where   $\tl H$ is the continuous function on $\{(\xi,\th):0\leq
\xi\leq
{\th\over 2}\leq {\pi\over 2}\}$, zero near the boundary of the
domain, and
given in the interior  by :
$${\tl H}(\xi,\th) ={{\cos\xi.
H\bigl(t(\xi,\th),\th\bigr)}\over{\sqrt{
\sin^2{\th\over 2} -   \sin^2 \xi} } } $$

Thus $\xi\mapsto\int_{2\xi}^\pi {\tl H}(\xi,\th)\,d\th$ is also
continuous.
Therefore, again by the Riemann-Lebesgue lemma, it follows that
$\lim_{n\to\infty}J_n=0$.

Putting all this together, we obtain equation $(2.5)$. $\bull$

Combining Lemma 2.17 and Proposition 2.19 we obtain (after uniformly
approximating   continuous $F$ by smooth ones) :

{\it  2.20. Proposition.} If $F$ is continuous on $G^2$  then :
$$\lim_{T\downarrow 0}{1\over {Z^I_T}}\int_{G^2} F(a,b)Q_T(
b^{-1}a^{-1}ba)\,da
db =\int_{K\times K} {\ov F}(k,k')\,dk.dk',$$
where $dk$ and $dk'$ are the Haar measure of unit total mass on  $K$,
and ${\ov
F}(a,b)$ is  the `average'  $\int_{SU(2)} F(gag^{-1},gbg^{-1})\,dg$.
(If
$F(gag^{-1},gbg^{-1})=F(a,b)$ for every $g\in SU(2)$, which is the
case of main
interest,  then ${\ov F}=F$.)

\vskip .25in

\noindent{\bf  3.  The  Symplectic Structure on Flat Connections}.
\vskip .10in

{\it 3.1. Tangent vectors to $\ca$} : Let $\ca$ be the space of all
connections
on a principal $\ovg-$bundle $\pi:P\to\S$. The space of  tangent
vectors to
$\ca$ is naturally identified as the space of $\ug-$valued $1-$forms
$A$ on $P$
satisfying :(i) $A\bigl((R_g)_*X\bigr)=Ad(g^{-1})A(X)$ for every
$g\in \ovg$,
$p\in P$ (we have written $R_gp$ to denote the action of $g\in\ovg$
on $p\in P$
arising from the principal $\ovg-$bundle structure of $P$), and $X\in
T_pP$,
and (ii) $A(Y)=0$ whenever $\pi_*Y=0$.

Recall that $\ug$ is equipped with an $Ad-$invariant metric
$\la\cdot,\cdot\ra_\ug$. If $\xi$ and $\eta$ are $\ug-$valued
$1-$forms on a
space we denote by $\la\xi\wedge\eta\ra$ the $2-$form defined by :
$$\la\xi\wedge\eta\ra(X,Y) =\la\xi (X),\eta(Y)\ra_\ug - \la\xi
(Y),\eta(X)\ra_\ug.$$

{\it 3.2. The symplectic form on $\ca$} : If $A^{(1)}$ and $A^{(2)}$
are
tangent vectors to $\ca$ then, as is easily verifiable,  $\la
A^{(1)}\wedge
A^{(2)}\ra$ is $\pi^*$ of a smooth $2-$form on  $\S$; the integral of
this
$2-$form over  (the oriented surface)  $\S$ will be denoted
$\Th(A^{(1)},A^{(2)})$. Thus $\Th$ itself is a $2-$form on $\ca$. In
fact,  it
is a symplectic form.

{\it 3.3. Notation $\O^\o$ }({\it curvature}){\it  and $\ca^0$} ({\it
flat
connections}) : Recall that the curvature of $\o$ is the $\ug-$valued
$2-$form
  $\O^\o$ specified  by
 $\O^\o(X,Y)=d\o(X,Y) +[\o(X),\o(Y)]$, for every $X,Y\in T_pP$ and
every $p\in
P$.
We denote by $\ca^0$ the space of  {\it flat connections} on $P$,
i.e. those
which have zero curvature.

{\it 3.4. Symplectic form on $\ca^0$ and $\cc^0$} : Let $\cg$ be the
group of
all bundle automorphisms of $P$ which cover the identity on $\S$,
i.e. $\cg$
consists of diffeomorphisms $\phi:P\to P$ for which
$\phi(pg)=\phi(p)g$ for
every $p\in P$ and $g\in \ovg$, and $\pi\phi=\pi$. Then $\cg$ acts on
$\ca$ by
pullbacks of connections : $(\phi,\o)\mapsto \phi^*\o$. The
symplectic form
$\Th$ induces a ``symplectic form''  ${\ov\Th}$ on the quotient space
$\cc^0={\ca^0}/\cg$ in the sense that if $A^{(1)}$ and $\As$ are
vectors in
$\ca$ tangent to $\ca^0$ (i.e. they are tangent to paths in $\ca^0$)
then :
$$\Th(\Af ,\As) =\Th(\phi^*\Af ,\phi^*\As)$$
for every $\phi\in\cg$.

Let $\S$ be the torus, as before, and recall from subsection 2.1  the
two basic
loops $S_1$ and $S_2$  which generate  the fundamental group
$\pi_1(\S,m)$
subject to the relation ${\ov S}_2{\ov S}_1S_2S_1=1$, in homotopy.
Recall
(Notation 2.4)  that  principal $\ovg-$bundles over $\S$ are
classified by
$h\in ker(G\to\ovg)$.

{\it 3.5. Notation}  ($\cf$, $\cf'$, $\cf_{sing}$, $\cf'/\ovg$)  :
Recall
(Notation 2.3) that $G=SU(2)$, $\ovg$ is either $SU(2)$ or $SO(3) =
SU(2)/\{\pm
I\}$, and $G\to \ovg :x\mapsto \ovx$ is the covering map.  Let :
$$\cf=\{(\ova, \ovb)\in \ovg^2 :  b^{-1}a^{-1}ba=h^{-1}\}
\eqno(3.1)$$

The group $\ov G$ acts on $\cf$ by : $\bigl(g,(x,y)\bigr)\mapsto
(gxg^{-1},
gyg^{-1})$.

Thus we have a quotient  $p : \cf\mapsto \cf/\ovg : (a,b)\mapsto
[a,b]$.

 If $h=I$ (i.e. the bundle $P$ is trivial) then, with $Z(\ovg)$ being
the
center of $\ovg$, we write
 $$\cf=\cf'  \cup  \cf_{sing},$$
 where $\cf_{sing}=  Z(\ovg)\times Z(\ovg)$, and
$\cf'=\cf\setminus\cf_{sing}$.
 If $h=-I$, we  set  $\cf' =\cf$. $\bull$

{\it 3.6. Fact / Notation} ($\Th$, $\ov\Th$) :  The map
$\ca^0/\cg_m\to\cf :
[\o]\mapsto  \bigl(g_u(S_1;\o),g_u(S_2;\o)\bigr)$ is a  bijection
(see  2.5
for $\cg_m$), and induces a bijection :

$$\cc^0\to \cf/\ovg : [\o]\mapsto [  g_u(S_1;\o),g_u(S_2;\o) ]
\eqno(3.2)$$

Using these bijections it is possible  (as described in Theorem 3.7
below) to
transfer the form $\Th$ to a $2-$form on $\cf'$, which descends to a
$2-$form
on $\cf'/\ovg$. In general, $\cf$ and $\cf/\ovg$ are not smooth
manifolds and
so care needs to be taken in working with ``$2-$forms'' on these
spaces. For
the case we are interested in, where $\S$ is the torus and $G=SU(2)$,
we will
describe the structure of $\cf$ and $\cf/\ovg$ and the corresponding
$2-$form
explicitly.

{\it 3.7. Theorem } ([KS 1]).  Let   $(\ova,\ovb)\in\cf$, and let
$\Af,\As,\Bf,\Bs\in\ug$ be such that  $(\ova\Af,\ovb\Bf) $  and
$(\ova\As,\ovb\Bs)$ are tangent to   $\cf$  at $(\ova,\ovb)$; i.e
they are the
initial tangent vectors of   smooth paths in $\ovg^2$ lying on $\cf$
and
passing through $(\ova,\ovb)$ initially.  Then each of these paths
corresponds
to a   path $\e\mapsto\o_\e$ in $\ca^0$ such that $(\e,x)\mapsto
\o_\e|_x$ is
smooth  (here $x\in P$). Moreover,  $\Th$ evaluated on the `tangent
vectors'
(i.e. the corresponding $1-$forms  ${{\partial \o_\e}\over
{\partial\e}}\big|_{\e=0}$) to these   paths in $\ca^0$ equals

$$\Th_{(\ova,\ovb)} \Bigl((\Af,\Bf), (\As,\Bs)\Bigr) \buildrel\rm
def\over =
{1\over 2}\Bigl\{\la \Af,\Bs\ra_\ug -\la
\As,\Bf\ra_\ug\Bigr\}\eqno(3.3)$$

 {\it 3.8. Theorem.}  Suppose $h=I$ (i.e. the bundle $P$ is trivial).
Let
$L$ be a  maximal torus in $\ovg$, to be definite :  $L=\ov K$; the
corresponding Weyl group is $W=N(L)/L=\{L,aL\}$, where $a\in
N(L)\setminus L$.
Let $L'=L\setminus Z(\ovg)$, and let $(L\times L)'=(L\times L')\cup
(L'\times
L)$. We consider the action of $W$ on $ (L\times L)'\times \ovg/L$
specified by
$aL.(l_1,l_2,gL) = (a l_1a^{-1}, al_2a^{-1},ga^{-1}L)$; the image of
$(l_1,l_2,xL)$ in the quotient $\bigl((L\times L)'\times
\ovg/L\bigr)/W$ will
be denoted  $[l_1,l_2,xL]_W$. Then :

(i) $\ovg$ acts smoothly on $\lpw$ by $g.[l_1,l_2,xL]_W =
[l_1,l_2,gxL]_W$;

(ii) $\cf'$ is a connected smooth   $4-$dimensional  submanifold of
$\ovg^2$,
and the map $\Psi :\lpw\to\cf' : [l_1,l_2,gL]_W \mapsto
(gl_1g^{-1},gl_2g^{-1})$ is a well-defined $\ovg-$equivariant
diffeomorphism;

(iii) $p' :\lpw\to (L \times L )'/W :[l_1,l_2,gL]_W \mapsto
[l_1,l_2]_W$ is a
smooth  fiber bundle over the smooth manifold $(L\times L)'/W$ with
fiber
$\ovg/L$ and  structure group $W$;

(iv) the map  $\psi :(L\times L)'/W\to \cf'/\ovg : [l_1,l_2]_W
\mapsto
[l_1,l_2]$ is a well-defined homeomorphism;

(v) the map $p :\cf'\to  \cf'/\ovg$ is a smooth submersion if and
only if
$\cf'/\ovg$ is equipped with the differentiable structure making
$\psi$ a
diffeomorphism;

(vi) $p\Psi =\psi p'$, and thus, by transferring  smooth  bundle
structure by
means of $\Psi$ and $\psi$,  $p:\cf'\to\cf'/\ovg$ is a smooth
fiber  bundle
with fiber $\ovg/L$ and  structure group $W$, and $(\Psi,\psi)$ is an
isomorphism of  such  bundles.

{\it Proof } :  (i) The mapping $\ovg\times (\lp)\to \lp :
\Bigl(g,(l_1,l_2,xL)\Bigr) \mapsto (l_1, l_2, gxL)$ is smooth and
$W-$equivariant. Since $\lp$ is a cover of $\lpw$, the latter
equipped with the
smooth structure making the covering map a local diffeomorphism, it
follows
that the induced action of $\ovg$ on $\lpw$ is smooth.

(ii)  The mapping $\Psi_L :L\times L\times \ovg/L\to \ovg^2  :
(l_1,l_2,gL)\mapsto (gl_1g^{-1},gl_2g^{-1})$  is  a well-defined
$\ovg-$equivariant  smooth map with image $\cf$,   and   maps $
(L\times
L)'\times \ovg/L$   onto $\cf'$. Its derivative at $(l_1,l_2,gL)\in
L\times
L\times \ovg/L$ is specified by the  map  (wherein $\underline l$ is
the Lie
algebra of $L$)

$$\underline l\times \underline l\times{\underline l}^\perp\to
\Psi_L(l_1,l_2,gL)^{-1}T_{\Psi_L(l_1,l_2,gL)}\ovg^2 $$
$$(A,B,X)\mapsto \Bigl(Ad(g)\{A +\bigl(Ad( l_1^{-1}) -1\bigr)X\},
Ad(g)\{B
+\bigl(Ad( l_2^{-1}) -1\bigr)X\}\Bigr).$$

  Examination of  this shows that $\Psi_L$  on $(L \times L)'\times
\ovg/L$ is
an immersion.

We show  that $\Psi_L$  restricted to   $ (L\times L)'\times
\ovg/L\to \cf' $
is a closed map.  Let $C$ be a relatively closed subset of $ (L\times
L)'\times
\ovg/L$, and let $\ov C$ be its closure in  $L\times L\times \ovg/L$.
Then $
{\ov C}\cap \bigl( (L\times L)'\times \ovg/L\bigr)=C$.  Moreover,
$\Psi_L^{-1}(\cf_{sing})=\{(L\times L)\setminus (L\times L)'\}\times
\ovg/L$.
It follows that $\Psi_L(C)=\Psi_L({\ov C})\cap \cf'$. Thus since
$\Psi_L({\ov
C})$ is closed (being compact) in $\ovg^2$, $\Psi_L(C)$ is closed in
$\cf'$.

  The map  ${ \Psi}:( (L\times L)'\times \ovg/L)/W\to\cf'$ is a
continuous
bijection. Since  $\Psi_L$ takes relatively closed subsets of  $
(L\times
L)'\times \ovg/L$ into  relatively closed subsets of   $ \cf'$, ${
\Psi}$ is
also a closed map. Therefore $ {\Psi}$ is   a homeomorphism.    Since
$\Psi_L$
is an immersion, it follows that  $ \Psi$ is  smooth  and is an
immersion.
Since $ \Psi$  is   a homeomorphism onto its image $\cf'$, and $
\Psi$ is also
an immersion, $\cf'$ is a submanifold of $ \ovg^2$ and $\Psi$ is a
diffeomorphism.
{}From   $\dim \bigl( (L\times L)'\times  \ovg/L\bigr)=4$, we
conclude that
   $\dim\cf'=4$.  Since $\Psi_L$  is $ \ovg-$equivariant, so is
$\Psi$.

(iii)    The smooth manifold $(L\times L)'$  is a  two-fold cover of
$(L\times
L)'/W$, and so the latter  has the smooth manifold structure induced
from that
of $(L\times L)'$. If $(l_1,l_2)\in (L\times L)'$ we can choose a
neighborhood
$U$ of $[l_1,l_2]_W$ in $(L\times L)'/W$ which is covered once by a
neighborhood $\tl U$ of $(l_1,l_2)\in (L\times L)'$. Consider $ f :
U\times
\ovg/L \to \lpw : ([j_1,j_2]_W, xL) \mapsto [j'_1, j'_2, xL]_W$,
where
$(j'_1,j'_2)\in \tl U$ covers $(j_1,j_2)\in U$. The map  $f$ is a
smooth
$\ovg-$equivariant  mapping onto $(p')^{-1}(U)$. Moreover, $f^{-1}$
is also
smooth since it is obtained by means  of  the smooth map $\lp\to
(L\times L)'/W
:(l_1,l_2,xL)\mapsto [l_1,l_2]_W$. Taking the maps $f$ as local
trivializations,  $p'$   becomes a smooth bundle with fiber $\ovg/L$.
Now  $W$
acts on $\ovg/L$ by : $(aL,x L ) \mapsto xa^{-1}L)$.  If $f_1$ and
$f_2$ are
local trivializations of the type $f$, defined on the same domain,
then
$f^{-1}_2f_1$ is smooth  and $f_2^{-1}f_1 ([l_1,l_2]_W,xL)$ equals
either
itself or $([l_1,l_2]_W,xaL)$. Thus  the bundle specified by $p'$ has
structure
group $W$.

(iv) It is clear that $\psi$ is well-defined, and that $\psi p'
=p\Psi$. Since
 $p$ is open, and since $p'$ has continuous local sections, it
follows that
$\psi$ is   continuous.  We show now that  $\psi$ is injective.
Suppose
$[l_1,l_2] =[l_3,l_4]$  in $\cf/\ovg$. Then there is a $g\in \ovg$,
for which
$gl_1g^{-1}=l_3$ and $gl_2g^{-1}=l_4$.  Since $l_1$ or $l_2$ is in $
L'$, it
follows from the structure of the elements of $ L=\ov K$, that either
$g\in  L$
or that $g=\ov h$ for some $h\in G$ of the form
$ \left(\matrix{  0 &    \b\cr
-{\ov\b}&   0\cr}\right)$
with  $|\b|=1$.  In either case,  $g\in N( L)$. Therefore,
$[l_1,l_2]_W=[l_3,l_4]_W$. So $\psi$ is one-to-one. Since $\Psi$ is
closed and
$p$ is closed (this is a general fact about quotients of Hausdorff
spaces by
compact groups) it follows that $\psi$ is closed. Therefore $\psi$ is
a
homeomorphism.

(v) If $p$ is a smooth submersion,  thereby having smooth local
sections
$p_{loc}^{-1}$, the map $\psi^{-1}$, being expressible locally as
$p'\Psi^{-1}p_{loc}^{-1}$, is also smooth.
Conversely, if $\psi $ is a diffeomorphism then  $p=\psi p'
\Psi^{-1}$ is a
smooth submersion.

(vi)  It is readily verified that $p\Psi =\psi p'$.   $\bull$

 {\it  3.9. Remark.}  The bundle $p':\bigl((L\times
L)'\times\ovg/L\bigr)/W\to
(L\times L)'/W$ is, as is seen directly  from the definition, the
bundle with
fiber  $\ovg/L$ associated to the principal $W-$bundle $(L\times
L)'\to
(L\times L)'/W : (l_1,l_2)\mapsto [l_1,l_2]_W$;  for this, note  that
$W$ acts
on the left on $\ovg/L$ by : $yL.xL= xy^{-1}L$ (for every $yL\in
W=N(L)/L$, and
$xL\in \ovg/L$). Since $(L\times L)'\to (L\times L)'/W$ is
non-trivial (because
$(L\times L)'$ is connected and $W$ is discrete  and has  more than
one
element), it follows that $p'$ is non-trivial. Hence {\it  the bundle
$p
:\cf'\to \cf'/\ovg$ is non-trivial}, and is a fiber  bundle, with
fiber
$\ovg/L$,  associated to the principal $W-$bundle  $(L\times L)'\to
(L\times
L)'/W$. $\bull$

{\it   3.10. Proposition}. Let  $(K\times K)'=(K\times K)\setminus
\bigl(Z(G)\times Z(G)\bigr)$, and let $ \rho$ denote the composite of
the
covering projection $(K\times K)'\to (L\times L)'$ with the covering
projection
$(L\times L)'\to (L\times L)'/W$.  Recall    from Theorem 3.8(iv)
the
diffeomorphism  $\psi :(L\times L)'/W\to \cf'/\ovg$.  Then the  form
$(\psi\rho)^*\ov\Th$ on $ (K\times  K)'$ is given by :

$$(\psi\rho)^*{\ov\Th}\Bigl((X_1,Y_1), (X_2,Y_2)\Bigr) = {1\over 2}
\Bigl( \la
X_1,Y_2\ra_\ug -\la X_2,Y_1\ra_\ug\Bigr).$$

In  particular,  the corresponding volume measure
$|(\psi\rho)^*\ov\Th|$ is
the restriction to $K'\times  K'$ of a Haar measure on $ K\times K$.
More
precisely, if $f$ is a bounded measurable function on $ K\times  K$
such that
$f(wx,wy)=f(x,y)$, for all $(w,x,y)\in W_K\times K^2$, wherein $W_K$
is the
Weyl group of $K$, then :

$$\int_{(K \times K)'} f \,d| (\psi\rho)^*{\ov\Th}| = {1\over {2}} |
K|^2\int_{
K\times  K} f(k,k')\,dkdk' ,$$

where $dk$ and $dk'$ are the Haar measure on $ K$ of unit total mass,
and $|
K|$ is the `volume' of $ K$ as measured by the restriction of the
metric
$\la\cdot,\cdot\ra_\ug$ to $K$.

{\it Proof } :  The expression for $(\psi\rho)^*\ov\Th$  follows from
Theorem
3.7 and the definitions of $\psi$ and $\rho$.  Since
$(\psi\rho)^*\ov\Th$ is
manifestly  a  translation invariant non-degenerate $2-$form on
$K\times K$,
the corresponding measure is a  constant multiple of the unit-mass
Haar measure
$dkdk'$. The factor  ${1\over {2}}|K|^2 $ is then obtained by
comparing the two
sides of the above equation in the case $f=1$. $ \bull$

{\it   3.11. Lemma}. If $f$ is a continuous  function on $\ovg\times
\ovg$
which is invariant under the conjugation action of $\ovg$, then :

$$\lim_{T\downarrow 0} {1\over {Z^I_T}} \int_{G\times G} f(\ovx,\ovy)
Q_T(
y^{-1}x^{-1}yx)\,dxdy = {1\over
{\hbox{vol}_{\ov\Th}\Bigl(\cf'/\ovg\Bigr)   }}
\int_{\cf'/\ovg} {\tl  f}\,d|\ov\Th|$$
where on the right  we have written ${\tl f}$   to denote the
function on
$\cf'/\ovg$ induced by the conjugation-invariant  function $f$,
$|\ov\Th|$ is
the measure corresponding to the  `volume form' $\ov\Th$, and (with
$|K|$ the
`volume' of  $K$ as measured by the metric  induced from $G$) :

$$\hbox{vol}_{\ov\Th}\Bigl(\cf'/\ovg\Bigr) = \cases{ {1\over 4}|K|^2
&  if
$\ovg=SU(2)$;\cr
 {1\over 8 }|K|^2, & if $\ovg=SO(3)$. \cr}$$

{\it Proof } :  This follows   from  Proposition  3.10  and
Proposition 2.20,
and the observation  that $\psi\rho$ is a two-to-one cover  of
$\cf'/\ovg$ in
case $\ovg=G=SU(2)$, while if    $\ovg=SO(3)$ then $\psi\rho$ is a
four-to-one
cover.  The factor  $\hbox{vol}_{\ov\Th}\Bigl(\cf'/\ovg\Bigr)$ is the
volume
of    $\cf'/\ovg$ as measured by the symplectic form $\ov\Th$.
$\bull$

{\it   3.12. Theorem. }  Let $\mu_T$ denote the YM measure, described
formally
by  $d\mu_T(\o)= {1\over {Z_T}} e^{-S_{YM}(\o)/T} [{\cal D}\o]$,  on
the
`moduli space'  $\cc$ of (generalized) connections on a principal
$SU(2)-$bundle or the trivial $SO(3)-$bundle, over the torus $\S$.
Let
$C_1,...,C_k$ be loops on $\S$ (which are composites of
$1-$simplices of a
triangulation as in  subsection  2.2).  Then :

$$\lim_{T\downarrow 0} \int_{\cc} f\bigl(g(C_1; \o ),...,g(
C_k;\o)\bigr)
\,d\mu_T(\o) ={1\over {\hbox{vol}_{\ov\Th}\Bigl(\cf'/\ovg\Bigr)
}}\int_{\cf'/\ovg}f\bigl(g(C_1;\o),...,g( C_k;\o )\bigr)\,
d|\ov\Th|(\o),$$
wherein $\ovg$ is $SU(2)$ or $SO(3)$, as before, and we have
identified the
space of flat connections with $\cf$.

{\it Proof } : This follows directly from Lemma 2.14, Lemma  3.11,
and equation
$(2.1')$ in Remark 2.12. $\bull$

{\it  3.13. Lemma. }   Suppose $a,b\in SU(2)$, and $
b^{-1}a^{-1}ba=-I$. Then
there is a $g\in SU(2)$ with

$$gag^{-1} =   \left(\matrix{  i & 0\,   \cr
 0 &   -i\cr}\right)$$

and

$$gbg^{-1} =  \left(\matrix{  0 & 1\,  \cr
-1 &   0\cr}\right).$$
An element  $x\in SU(2)$ commutes with both $a$ and $b$ if and only
if  $x
\in\{\pm I\}$.

{\it Proof } : Without loss of generality, assume that $a  = k_\phi$,
with
$\phi\in [0,\pi]$. Then, calculating $b^{-1}a^{-1}ba$, we see by
direct
computation that this commutator equals $-I$ if and only if $b$ is of
the form
$ \left(\matrix{  0 & e^{i\psi}\,  \cr
-{e^{-i\psi}}&   0\cr}\right)$ and $\phi$ is $\pi\over 2$.  Now let
$g=k_{
\psi/2}^{-1}$. Then $gag^{-1}=a$, and, as calculation shows,
$gbg^{-1} =
\left(\matrix{  0 & 1\,  \cr
-1 &   0\cr}\right)$.

If  $x\in G$ commutes with $ \left(\matrix{  i & 0\,   \cr
 0 &   -i\cr}\right)$  then $x\in K$, i.e. $x= \left(\matrix{  \a &
0\,  \cr
0 &   {\ov\a}\cr}\right)$, for some $\a$ with $|\a|=1$.  Direct
calculation
shows that  if  this $x$ commutes with $ \left(\matrix{  0 & 1\,  \cr
-1 &   0\cr}\right)$  then  $\a^2=1$, and so  $x=\pm I.\quad$
$\bull$

Combining Fact 3.5 and Lemma  3.13  we obtain :

{\it   3.14. Proposition. } There is, up to bundle automorphism,
exactly  one
flat connection on the non-trivial $SO(3)-$bundle over the torus.
$\bull$

{\it  3.15. Theorem.}  Let $\mu_T$ be the Yang-Mills measure  for the
non-trivial $SO(3)-$bundle over $\S$.  Then :

$$\lim_{T\downarrow 0}{1\over {Z^{-I}_T} }  \int_\cc
f\Bigl(g(C_1;\o),...,g(
C_k;\o)\Bigr)\,d\mu_T(\o) =f\Bigl(g(C_1;\o^0),...,g(
C_k;\o^0)\Bigr)$$

for any continuous $SO(3)-$invariant function $f$ on $SO(3)^k$ (i.e.
$f(gx_1g^{-1},...,gx_kg^{-1})=f(x_1,...,x_k)$ for every
$g,x_1,...,x_k\in
SO(3)$) and  loops $C_1,..., C_k$  on $\S$ (satisfying the conditions
of  2.2),
and $\o^0$  is  `the' flat connection on the bundle.

{\it Proof } : As in the proof of  Theorem   3.12, it will suffice to
show
that for every continuous $SU(2)-$invariant function  $F$ on
$SU(2)\times
SU(2)$ :
$$\lim_{T\downarrow 0}{1\over {Z^{-I}_T}} \int_{SU(2)\times SU(2)}
F(x,y)
Q_T(y^{-1}x^{-1}yxh)\,dxdy =F(a,b) \eqno(3.4)$$

wherein $h=-I$, and $a = \left(\matrix{  i & 0\,   \cr
 0 &   -i\cr}\right)$ and $b= \left(\matrix{  0 & 1\,  \cr
-1 &   0\cr}\right).$

Let $M:G\times G\to G :(x,y)\mapsto y^{-1}x^{-1}yx$.  We write
$M'_{(x,y)}:\ug\times\ug\to\ug$ for the `translated  derivative' of
$M$ at
$(x,y)\in G^2$; i.e.
$$M'_{(x,y)}(X,Y) \buildrel\rm def\over =
M(x,y)^{-1}dM_{(x,y)}(xX,yY).$$
Thus :

$$M'_{(x,y)}(X,Y) =\{  1 - Ad(x^{-1}y^{-1}x)\}X + \{Ad(x^{-1}) -
Ad(x^{-1}y^{-1}xy)\}Y\eqno(3.5)$$

So  if  $(x,y)\in M^{-1}(z)$, wherein $z\neq I$,  then $M'_{(x,y)}$
is
surjective; for, as some algebraic manipulation  shows,  any  $Z$
orthogonal to
the image of $M'_{(x,y)}$  satsifies both $Ad(x)Z=Z$ and $Ad(y)Z=Z$,
which,
since $x$ and $y$ do not commute, imply that $Z$ is zero (recall that
we are
working in the Lie algebra of $SU(2)$).  Therefore, for every $z\in
SU(2)\setminus \{I\}$,     $M^{-1}(z)$ is a smooth $3-$dimensional
closed
submanifold of  $SU(2)\times SU(2)$.

 For the sake of  computational convenience it will also be
convenient to
assume that the metric $\la\cdot,\cdot\ra_\ug$ on $\ug$ is scaled so
that $G$
has unit  total volume; in this case, the Haar measure $dx$ is the
same as  the
Riemannian volume measure (this rescaling does not alter either  side
of
equation $(3.4)$).  We  claim  that :

$$\int_{G^2} F(x,y) Q_T(y^{-1}x^{-1}yxh)\,dxdy =\int_{G\setminus
\{I\}} dz.
Q_T(zh )\Bigl[ \int_{M^{-1}(z)} {{F(x,y)}\over {J(x,y)}}\,dv_z(x,y)
\Bigr]
\eqno(3.6)$$
where $dv_z$ is the   Riemannian volume measure  on $M^{-1}(z)$
corresponding
to the metric induced by means of  $\la\cdot,\cdot\ra_\ug$, and
$J(x,y)$ is the
Jacobian factor  $ |det \{M'_{(x,y)}(M'_{(x,y) })^*\}|^{1/2}$. As we
have seen
above, $J>0$ off $M^{-1}(I)$.  Since $M^{-1}(I)$ has measure zero (by
Fubini's
theorem, since for every $x\in G\setminus Z(G)$, the set $\{y\in G:
M(x,y)=I\}$, being a two-dimensional torus in $G$, has Haar measure
zero) we
can use  a monotone limit argument  to see that it suffices to prove
$(3.6)$
under the assumption that $F$ vanishes in a neighborhood  $N$ of
$M^{-1}(I)$
(actually this, in essence,  is the only case we  really need for our
purposes).   Consider any $(x,y)\in G^2\setminus N$, and let   $
M(x,y)=z\neq
I$; then by the inverse function theorem,  there is a neighborhood
$W(\subset
G^2\setminus N)$ of $(x,y)$  in $G^2$ and a diffeomorphism   $\Phi :
W\to
V\times U$, where $V$ is a neighborhood  of $(x,y)$ in $M^{-1}(z)$
and $U$ is a
neighborhood  of $z$ in $G$ such that  $\Phi $, on $W$,  is of the
form
$\bigl(\ast, M(\cdot,\cdot)\bigr)$.  To prove equation $(3.6)$ it
will be
sufficient  (by a partition of unity argument) to assume that $F$ has
support
in $W$.  However, for $F$ supported in $W$, equation $(3.6)$  is just
a
`change-of-variable' formula, and   $\int_{M^{-1}(z)} {{F(x,y)}\over
{J(x,y)}}\,dv_z(x,y)$  depends continuously  on $z\in
G\setminus\{I\}$. Thus
$(3.6)$ is proved for all continuous $F$, and the integrand
$\int_{M^{-1}(z)}
{{F(x,y)}\over {J(x,y)}}\,dv_z(x,y)$  depends continuously  on $z\in
G\setminus\{I\}$.

     Therefore, equation $(3.6)$ and Facts 2.9  imply  (by splitting
the
integral on the left below into a  sum of two integrals, one outside
a small
neighborhood  of $M^{-1}(I)$ and another, contributing zero  in the
limit,
over the small neighborhood) :

$$\lim_{T\downarrow 0} \int_{G^2} F(x,y) Q_T(y^{-1}x^{-1}yxh)\,dxdy
=\int_{M^{-1}(h^{-1})} {{F(x,y)}\over {J(x,y)}}\,dv_{h^{-1}}(x,y)
\eqno(3.7)$$

wherein $h=-I$.
Setting $F=1$, we see again that  $\lim_{T\downarrow 0}Z^{-I}_T$
exists and is
finite; dividing both sides of equation $(3.7)$ by this we obtain
equation
$(3.4)$  by means of  Lemma  3.13 .
$\bull$

{\it Another proof of  the second part of  Lemma 2.13 } : The aim is
to
calculate $\lim_{T\downarrow 0}Z_T^{-I}$. We will use the notation
and
conventions  introduced in the proof  of  Theorem  3.15 above. By
Lemma  3.13
and the observation that $J(x,y)=J(gxg^{-1},gyg^{-1})$ for every
$(g,x,y)\in
G^3$,  we see that the right hand side of  equation $(3.7)$ equals
$F(a,b)J(a,b)^{-1}\int_{M^{-1}(-I)} dv_{-I}(x,y)$.
 Thus
$$\lim_{T\downarrow 0} \int_{G^2}  Q_T(y^{-1}x^{-1}yxh)\,dxdy =
{1\over
{J(a,b)}}\hbox{vol}\Bigl(M^{-1}(-I)\Bigr),$$

where the volume   $\hbox{vol}\Bigl(M^{-1}(-I)\Bigr)$  is with
respect to
$dv_{-I}$.
By Lemma  3.13,  for $(a,b)\in M^{-1}(-I)$,  the smooth map $\Psi :
G\to
M^{-1}(-I) : x\mapsto (xax^{-1},xbx^{-1})$ is surjective and
two-to-one.  Let
$\Psi'_x :\ug\to \ug\times\ug :X\mapsto \Psi(x)^{-1}d\Psi_{x}(xX)$.
Then
calculation shows that $(\Psi'_x)^*(\Psi'_x) =M'_{(a,b)}(M'_{(a,b)
})^*$ (both
have the same diagonal matrix, with diagonal entries $(4,4,8)$,
relative to
the basis  $\Bigl\{ \left(\matrix{  i & 0\,   \cr
 0 &   -i\cr}\right),   \left(\matrix{  0 & 1\,  \cr
-1 &   0\cr}\right),   \left(\matrix{  0 & i\,  \cr
i &   0\cr}\right) \Bigr\}$ of $\ug$). Therefore,
$\hbox{vol}\Bigl(M^{-1}(-I)\Bigr)$ equals ${1\over 2}J(a,b)$.
Therefore,

$$\lim_{T\downarrow 0} \int_{G^2}  Q_T(y^{-1}x^{-1}yxh)\,dxdy =
{1\over
2}.\qquad\bull$$

{\bf Acknowledgement}. It is a pleasure to thank  Chris King for many
useful
conversations.  Part of  this work was done in July 1991 at Cornell
University,
and I thank Leonard Gross for arranging supporting for me at that
time through
NSF Grant  DMS-8922941.

 \vfill\eject
\RefHead

\References{BrtD}{T. Brocker and T. tom Dieck,
{\it Representations of Compact Lie Groups},
Springer-Verlag (1985)}
\References{Fo}{R. Forman, {\it Small volume
limits of 2-d Yang-Mills}, Commun. Math. Phys. {\bf 151},
39-52 (1993)}
\References{KS 1}{ C. King and A. Sengupta, {\it  An Explicit
Description of
the  Symplectic Structure   of  Moduli Spaces of Flat Connections},
preprint
(1994)}
\References{KS 2}{ C. King and A. Sengupta, {\it The Semiclassical
Limit  of the Two  Dimensional Quantum Yang-Mills Model}, preprint
(1994)}
\References{Se 2}{A. Sengupta, {\it The Semiclassical
Limit  for Gauge Theory on ${\rm S}^{2}$}, Commun. Math, Phys. {\bf
147},
191-197 (1992)}
\References{Se 3}{A. Sengupta, {\it Quantum Gauge
Theory on Compact Surfaces}, Ann. Phys. {\bf 221},
17-52 (1993)}
\References{Se 4}{A. Sengupta, {\it Gauge Theory
on Compact  Surfaces}, preprint (1993)}
\References{Waw}{A. Wawrzynczyk, {\it Group Representations and
Special
Functions}, D. Reidel Publishing Company (1984)}
\References{Wi1}{E. Witten, {\it On Quantum
Gauge Theories in Two Dimensions}, Commun. Math.
Phys. {\bf 141}, 153-209 (1991)}
\References{Wi 2}{E. Witten, {\it Two Dimensional
Quantum Gauge Theory Revisited}, J. Geom.
Phys. {\bf 9}, 303-368 (1992)}

\bye